\documentclass[conference]{IEEEtran}
\IEEEoverridecommandlockouts
\usepackage{cite}
\usepackage{amsmath,amssymb,amsfonts}
\usepackage{algorithmic}
\usepackage{graphicx}
\usepackage{textcomp}
\usepackage{xcolor}
\usepackage{hyperref}

\def\BibTeX{{\rm B\kern-.05em{\sc i\kern-.025em b}\kern-.08em
    T\kern-.1667em\lower.7ex\hbox{E}\kern-.125emX}}
\begin{document}

\title{Towards formally analyzed Cyber-Physical Systems
\thanks{
This work was partially supported by the ÚNKP-20-4-II. New National Excellence Program of the Ministry of Human Capacities and the European Commission and the Hungarian Authorities (NKFIH) through the Arrowhead Tools project (EU grant agreement No. 826452, – NKFIH grant 2019-2.1.3-NEMZ ECSEL-2019-00003.}
}

\author{\IEEEauthorblockN{1\textsuperscript{st} Richárd Szabó}
\IEEEauthorblockA{\textit{Dept. of Measurement and Information Systems} \\
\textit{Budapest University of Technology and Economics}\\
Budapest, Hungary \\
richard.szabo@edu.bme.hu}
\and
\IEEEauthorblockN{2\textsuperscript{nd} András Vörös}
\IEEEauthorblockA{\textit{Dept. of Measurement and Information Systems} \\
\textit{Budapest University of Technology and Economics}\\
Budapest, Hungary \\
voros.andras@vik.bme.hu}
}

\maketitle

\begin{abstract}
Cyber-physical systems (CPS) can be found everywhere: smart homes, autonomous vehicles, aircrafts, healthcare, agriculture and industrial production lines. 
CPSs are often critical, as system failure can cause serious damage to property and human lives.

Today's cyber-physical systems are extremely complex, heterogeneous systems: to be able to manage their complexity in a unified way, we need an infrastructure that ensures that our systems operate with the high reliability as intended. In addition to the infrastructure, we need to provide engineers a method to ensure system reliability at design time. The paradigm of model-driven design provides a toolkit supporting the design and analysis and by choosing the proper formalisms, the model-driven design approach allows us to validate our system at design time.
\end{abstract}


\section{Introduction}
Today’s cyber-physical systems (CPS) are extremely complex, heterogeneous systems that contain a wide variety of devices, from sensor networks through edge devices to cloud-based services.

Our work focuses on edge-based CPSs, where increased computing capacity allows us to perform more complex computational tasks directly close to the data source. In these distributed systems the challenge is the changeability of the system. Hardware devices can fail or become inaccessible, causing services to fall out, temporarily or even completely. 

To address this challenge, we need an infrastructure that ensures that our system operates with high reliability and fault-tolerance. Both the communication middleware and the containerization technologies play an important role in the operation of the infrastructure, as we can use these technologies to satisfy the strict extra-functional requirements.

In addition to the right infrastructure, we need to provide engineers with the ability to ensure the dependability of the system at design time. The paradigm of model-based development approach provides a toolkit for this. By choosing the right formalisms, the model-based development approach allows us to verify our system at design time. 

In this paper we propose an architecture using fault-tolerant technologies to operate critical CPS, a model-based development approach which enables engineers to focus on specific aspects of the development of CPS and verification workflow integrating existing verification tools.

\section{Background}
Cyber-physical systems (CPS) are complex systems integrating the physical world with computerized infrastructures. CPS consists of sensors, embedded devices, edge and cloud services and components. Modern CPSs are often regarded as system-of-systems (SoS) based on the integration of subsystems. These subsystems are specialized, and the intelligence of the system originates in the integration of these subsystems\cite{nist-cps}.

In a system integrating many critical subsystems, the dependability of the system must be ensured from several aspects. First, the subsystems must operate in a safe and fault free way. Second, the communication between the integrated subsystems must be reliable and realtime, so every subsystem gets the required data instantly. Third, the runtime environment must be fault-tolerant, so the services provided by the subsystems are operated with high availability.

\subsection{Gamma Statechart Composition Framework}
Gamma is a statechart composition framework \cite{gamma} designed to model and verify component-based reactive systems and generate code from the models. The tool supports the hierarchical composition of the statecharts, which enables engineers to focus on subcomponents of the complex systems. 
The framework raises a new modeling layer over Yakindu \cite{yakindu} state machines to describe communication between state machines, making it suitable for modeling complex, hierarchical systems. The tool allows the user to formally verify the completed models using UPPAAL \cite{uppaal}. Finally, specific code can be generated from the models.

\subsection{Data Distribution Service}
The Object Magament Group's (OMG) Data Distribution Service (DDS)  \cite{omg-dds-spec} is a middleware targeting realtime distributed applications. DDS describes a Data-Centric Publish-Subscribe (DCPS) model, which enables simple integration between the distributed services and provides a reliable, robust communication.

DDS provides a virtual global data space where participants can publish and read data objects. The global data space is divided into domains and topics are defined in each domain. The topics are the main channel of communication, participant can subscribe to and publish to these topics. The domains designate planes in the data space, and although identical topics may appear in multiple domains, participants can only communicate with participants in the same domain.

Quality of Service (QoS) characteristics are a set of features that can be used to influence an aspect of how DDS service works. These are settings of the system that can be used to define that the system should work reliably, how participants can access the data objects. Moreover, it provides security features to the topics, or even behaviours that allow participants who joined the system later to access previously published data. 

\subsection{Kubernetes}

Kubernetes \cite{kubernetes} is a framework for automating deployment and managing the lifecycle of containerized applications. The framework is capable of deploying, scaling and restarting applications. It also provides means to securly store sensitive informations for our applications. With its affinity and antiaffinity labeling it is possible to define which device is capable to run which application, and which applications should not be deployed together. 

\subsection{Related Work}

Due to the importance of critical CPSs and the several aspects these systems can be analysed from, there are a vast number of works dealing with modeling approaches. In \cite{improving-cps-modeling} the importance of well defined models and appropriate tool chains are pointed out in the development of CPSs. In \cite{cps-security} a taxonomy is described to characterize the components of the CPS and a methodology to find security risks in it. In \cite{pvs-simulink} a methodology is provided and tool support is presented to formally verify and validate cooperating discrete and continuous CPS components. The components of CPSs can work independently, so in \cite{cps-autonomy} a modeling approach is decribed to model autonomous CPSs and to analyze their collaboration.

In our work presented in this paper we focus on a similar, yet different aspect of the CPS development. Our goal is to aid the development process and automate the deployment of the CPS, while formally verifying the functional correctness of the system.  

\section{Overview of the approach}

Our approach tries to aid the full engineering workflow, from the desinging process, throught the system verification to the deployment, while providing a fault-tolerant architecture and automating most of the work, so the engineer can focus on the system design.

\subsection{Architecture of critical CPS}
\begin{figure}[htbp]
    \centerline{
        \includegraphics[width=0.8\linewidth]{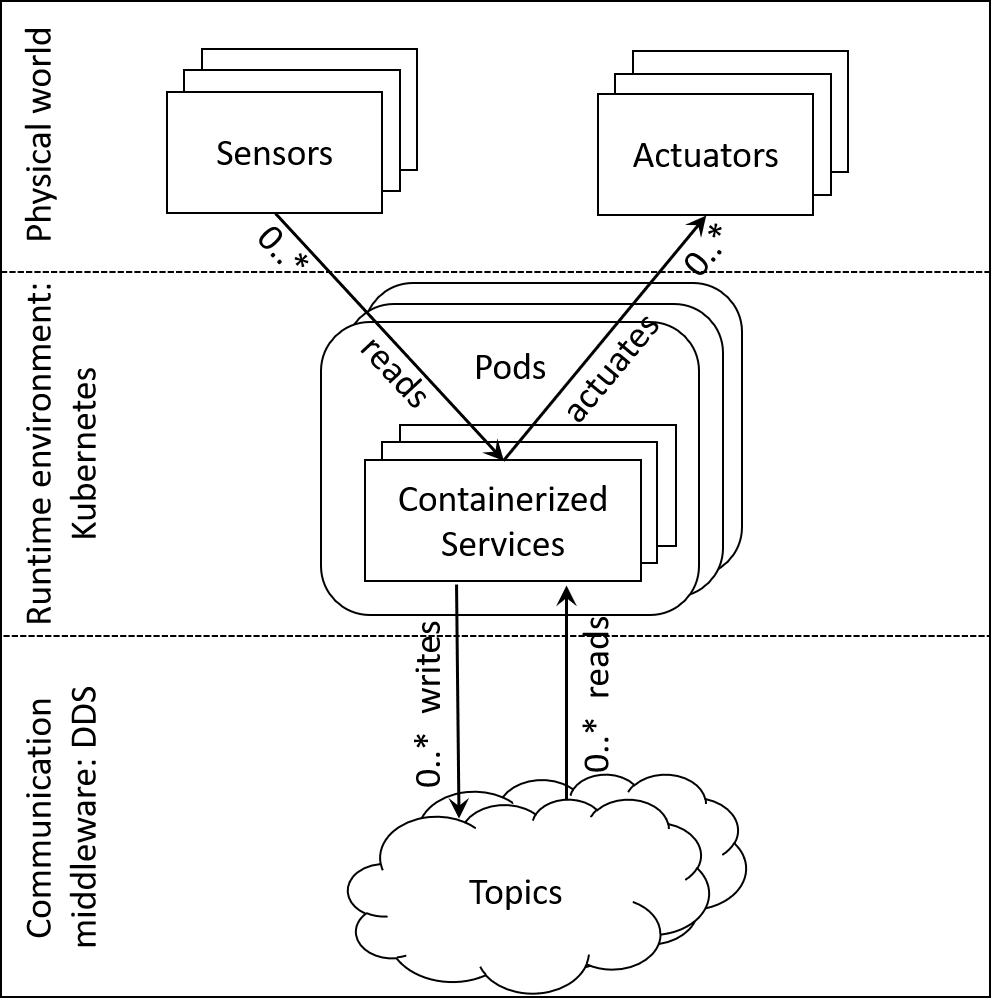}
    }
    \caption{Critical CPS architecture.}
    \label{fig:critical-cps-arch}
\end{figure}

Fig. \ref{fig:critical-cps-arch} shows the proposed architecture. 

The components of the system are deployed to hardware devices which have access to sensors and actuators, which can be used to interact with the physical world. 

The software of the components are run by Kubernetes, which can dynamically deploy components to hardware devices with suitable sensors and peripherals. In the event of a dropout of a hardware device, Kubernetes can redeploy the components to other suitable hardware devices, thus providing a fault-tolerant runtime environment.

The communication between the components is provided by the DDS middleware. Its dynamic discovery service provides a way for the components to communicate regardless of which device they are deployed to. With its history, DDS can provide previously published data to redeployed components, so they can continue the operation based on the last valid data, even if it was published after the dropout and before the redeploy.

\subsection{Modeling approach}

The modeling approach proposed in this paper captures three levels of abstraction. These levels of modeling provide an opportunity for the design engineer to focus on one aspect of the system under design at a time. The modeling levels are the following:
\begin{itemize}
	\item \textbf{High-level model}: At the top is the high-level model. This model gives a high-level description of the system. Following the SoS approach, the model provides hierarchical composition of the components and describes the basic capabilities, interfaces and connections of the subsystems in the system. 
	\item \textbf{Communication and behaviour model}: In the middle, we model the internal operation of our systems and the communication between them. This level consists of two kinds of models, the communication model and the behaviour model. The communication model can be used to describe the communication interfaces of the system and subsystems, and their QoS characteristics, while the behaviour model can be used to describe the internal operation of the subsystems. 
	\item \textbf{Deployment model}: At the bottom is the deployment model. The deployment model can be used to describe the physical components in the system and which physical components the subsystems should be installed on.
\end{itemize}

As the first step in facilitating the design workflow, we introduced model transformations that form a gateway between the modeling levels. The purpose of model transformations is to avoid the need to recreate relevant parts of the system at different levels of modeling, as they have already been designed at an earlier level, so they can be used in the design process of the next level. 

The second step is to generate executable files using the completed models. From the behaviour and communication model we generate the executable code, which is run in the containerized services, and from the deployment model we generate the deployment artifacts, which is used by the Kubernetes to run these services. The behaviour model is also used to verify the functional correctness of the system.

\begin{figure}[htbp]
\centerline{\includegraphics[width=0.8\linewidth]{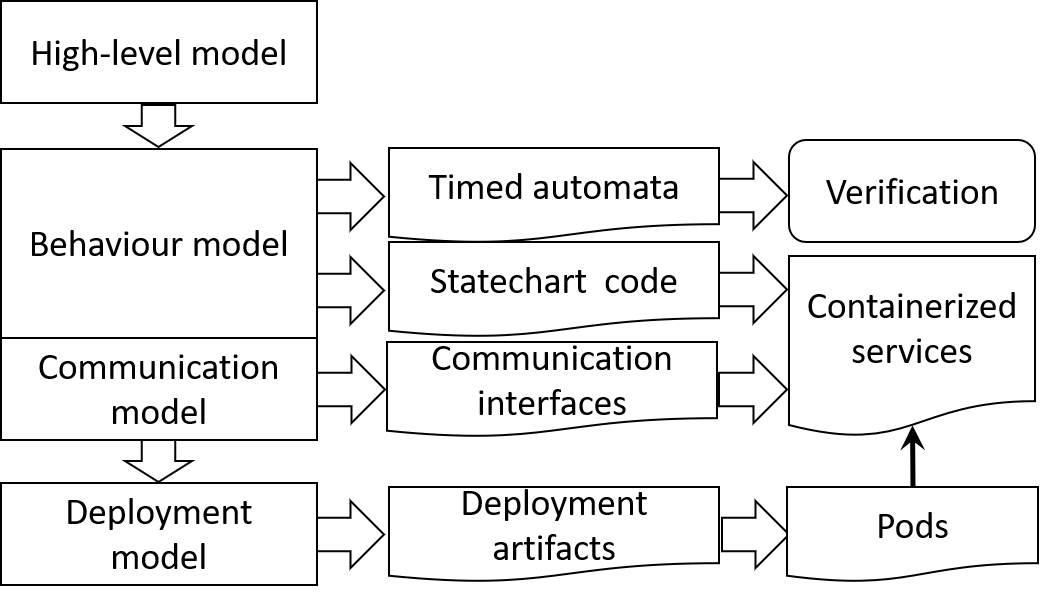}}
\caption{Overview of the model transformations and code generations.}
\label{fig:model-transformations}
\end{figure}

The introduced model transformations and code generations can be seen on Fig. \ref{fig:model-transformations}.

\subsection{High-level system description}
The high-level model is based on the Semantic Sensor Network (SSN)  \cite{ssn} ontology. The SSN ontology is designed to describe sensors and their observations, procedures processing these observations, and actuators that operate on the basis of the processed observations.

The purpose of the high-level model is to define system components and the dataflow between those components, which can later be used to generate the skeletons of the precise behaviour models.  

The high-level model is suitable for expressing system-subsystem relationships between the components, thereby supporting the hierarchical composition of the system. Every subsystem can have inputs and outputs. These have specified interfaces, which describe the type of data the given output provides or the given input expects. The system also has requirements that can be used to formulate expectations that must be taken into account during deployment, for example the component needs a temperature sensor, so it can only be deployed to a hardware device which has at least one temperature sensor. Finally, we can describe the properties of our subsystems. 

\subsection{Communication model}
The communication model describes the communication interfaces of the system services. The elements of the model correspond to the model elements of DDS.

This model captures the services provided by the components of the system. The main part of the model contains the elements that describe the communication participants belonging to the services, while the rest describes the QoS properties of the participants.

\subsection{Behaviour model}
To take advantage of the potential of a model-driven approach, we create behavioural models for the reactive components of our system. Behaviour models have formalisms that we can use to ensure the proper functioning of our system, thus ensuring the dependability of the system. 

For behaviour modeling, we chose an existing framework, the Gamma Statechart Composition Framework, which supports the modeling, verification, and generation of code from reactive components.

\subsubsection{State-based models}

A tool for describing the behaviour of reactive controllers is the statechart formalism. Statecharts are hierarchical state machines whose states can be refined into additional sub-states. This makes it possible to describe not only the simple components but also the behaviour of the entire system. The statechart formalism also provides a way to describe competing regions, which can be used to define several reactions to one event. This way we can further decompose the behaviour. 

\subsubsection{Verification}

For critical components, it is important to prove that the behaviour of the components are correct. There are several methods for proving correctness, one of which is model checking on timed automata.

Timed automata is a matematical formalism to describe the behaviour of the system, while statechart is a higher level formalism over the timed automata. Therefore if we have a statechart describing the behaviour of the system, we can transform the model to a timed automata.

Temporal logic expressions can be used to check a timed automata for requirements such that the automata is deadlock free or it will never enter a faulty state. 

\subsubsection{Behaviour modeling and verification workflow}

\begin{figure}[htbp]
\centerline{\includegraphics[width=0.7\linewidth]{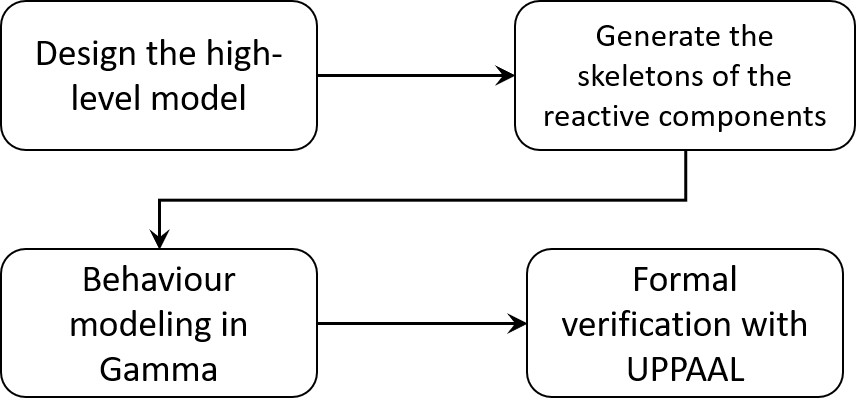}}
\caption{Behaviour modeling and verification workflow.}
\label{fig:behaviour-modeling-verification-workflow}
\end{figure}

In our approach we defined a workflow integrating the Gamma Framework as seen on Fig. \ref{fig:behaviour-modeling-verification-workflow}.

First, the engineer designs the high-level model, defining the components of the system, their interfaces and the relations between them. Second, using the information from the high-level model, the engineer generates the skeletons of the Gamma models. Third, the engineer designs the behaviour of the components. Fourth, using the Gamma Framework, the engineer generates the formally verifiable automata from the Gamma models. Finally, using the UPPAAL verification tool, the engineer formally verifies the models.

\subsection{Deployment model}

The deployment model describes the hardware components of the system and the needs of the services to be installed. The services to be installed can be divided into several basic types according to their needs, depending on whether they are stateless (their output depends only on their input) or stateful (their output depends on their input and their previous state). 

Another dedicated type is fault-tolerant services. Since the applications of CPS are critical, fault-tolerant services have distinguished role and are treated with particular care. These services include voters or other validation services.

An important aspect of service deployment is the relationship of services to each other. Certain services cannot be installed side by side, for example because they provide the same service and a voter decides the outcome of the service based on the results of the implementations. In this case, to ensure the independence of the instances, the instances must be installed on separate hardware components. The opposite may also be true, you may want to install certain services on a hardware device. 

Hardware devices have a type that describes the capabilities of a given hardware device. These capabilities can be resource parameters, or even sensors, peripherals. Hardware devices can be organized into groups, making it easier to manage devices that perform the same task. Hardware device groups means hardware redundancies, and services can be installed on any available unit in the group.

\section{Conclusion and future work}
In this paper we presented an architecture using fault-tolerant technologies to operate critical CPS, a model-based development approach which enables engineers to focus on specific aspects of the development of CPS and verifaction workflow integrating existing verification tools.

The presented modeling approach starts with the high-level model and leads to the deployment of the system, while integrating a verification framework to prove the functional correctness of the system.

Our future work focuses on the integration of other verification methods to verify other aspects of the system, automated integration testing and provide dependability analysis support.

\vspace{12pt}

\end{document}